\documentclass[a4paper,11pt]{article}
\pdfoutput=1 

\usepackage{jheppub} 

\usepackage[T1]{fontenc} 
\usepackage{comment}
\usepackage{soul}
\usepackage[textsize=footnotesize,textwidth=2.cm]{todonotes}
\usepackage{hyperref}

\def\p{\partial}
\newcommand{\ba}{\begin{array}}
\newcommand{\ea}{\end{array}}
\newcommand{\bi}{\begin{itemize}}
\newcommand{\ei}{\end{itemize}}
\newcommand{\bea}{\begin{eqnarray}}
\newcommand{\eea}{\end{eqnarray}}
\newcommand{\be}{\begin{equation}}
\newcommand{\ee}{\end{equation}}
\newcommand{\nn}{\nonumber}

\title{\boldmath Warped Conformal Symmetries of the Accelerating Kerr Black Hole}

\author{Jianfei Xu}
\date{\today}

\affiliation{Shing-Tung Yau Center and School of Mathematics, Southeast University, Nanjing, 210000, China}
\emailAdd{jfxu@seu.edu.cn}

\abstract{Recent studies on the holographic descriptions of Kerr black holes indicate that the conformal or the warped conformal symmetries are responsible for the Kerr black hole physics at both background and perturbation levels. In the present paper, we extend the validity of these studies to the case of accelerating Kerr black hole. By invoking a set of non-trivial diffeomorphisms near the horizon bifurcation surface of the accelerating Kerr black hole, the Dirac brackets among charges of the diffeomorphisms form the symmetry algebra of a warped CFT which consists of one Virasoro and one Kac-Moody algebra with central extensions. This provides the evidence for warped CFTs being possible holographic dual to accelerating Kerr black holes. The thermal entropy formula of the warped CFT fixed by modular parameters and vacuum charges reproduces the entropy of the rotating black hole with acceleration.}

\begin{document}
\maketitle
\flushbottom
\section{Introduction}
The holographic principle of gravity provides profound insights into both macroscopic and microscopic properties of black holes in general relativity. Stemming from the Anti de-Sitter (AdS)$/$conformal field theory (CFT) correspondence, the entropy of BPS black holes in string theory admits microscopic interpretations by state counting~\cite{Strominger:1996sh, Sen:2014aja}. The crucial mechanism that enables these descriptions is the conformal symmetry, which is applied from the AdS factor in the black hole near horizon geometry. To be specific, these descriptions rely on a concrete set up of the correspondence based on symmetry analysis in lower dimensions, which is known as the AdS$_3/$CFT$_2$~\cite{Brown:1986nw}. The infinite local symmetries in the 2D CFT help to determine the physical degrees of freedom in a bulk spacetime with AdS factor. For example, the asymptotic growth of states determines the entropy of black holes in string theory~\cite{Strominger:1997eq, Maldacena:1998bw}.

For spacetimes without AdS factors, the holographic principle still applies for many cases. Kerr black holes are typical celestial objects in universe. The holographic aspects of Kerr black holes have been widely explored both on the gravity side and on the field theory side. For extreme Kerr black holes, the near horizon region obtained by a scaling procedure shares the $SL(2, R)\times U(1)$ isometry~\cite{Bardeen:1999px}. Under specific boundary conditions, the asymptotic symmetry analysis shows that this scaling region has enhanced symmetries which are identical to the local symmetries of a chiral 2D CFT~\cite{Guica:2008mu, Matsuo:2009sj, Chen:2011wt}. This is known as the Kerr$/$CFT conjecture, supported by the evidence that the Cardy formula reproduces the extreme Kerr black hole entropy~\cite{Guica:2008mu, Matsuo:2009sj}, and thermal correlators of a 2D CFT match the scalar scattering amplitudes on an extreme Kerr background~\cite{Bredberg:2009pv}. For non-extreme Kerr black holes, the near horizon scaling region disappears and the spacetime geometry carries no explicit conformal symmetry. However, in this case, scalar perturbations with low frequency on a generic Kerr background can be shown to satisfy hypergeometric equations in the near region radial direction, which are manifestly invariant under global conformal transformations~\cite{Castro:2010fd}. This hidden conformal symmetry is intrinsic to the Kerr background and governs the dynamics of the scalar field on it, which is reflected from the facts that the black hole entropy is reproduced form the Cardy formula and the scalar scattering amplitudes coincide with thermal correlators of a 2D CFT~\cite{Castro:2010fd, Nian:2023dng}. Further more, a two copies of Virasoro algebra, which is the symmetry algebra belongs to a 2D CFT, can be explicitly built in the covariant phase space near the horizon bifurcation surface of a generic Kerr black hole~\cite{Haco:2018ske}.

In this paper, we will analysis the holographic descriptions of accelerating Kerr black holes in the covariant phase space. Black holes with acceleration parameters are also exact solutions to the Einstein equation with possible electromagnetic fields coupled. Such a solution is usually known as the C-metric which describes a pair of black holes constantly accelerating away from each other~\cite{Kinnersley:1970zw, Plebanski:1976gy, Dias:2002mi, Hong:2003gx, Hong:2004dm, Griffiths:2005se, Griffiths:2005qp, Podolsky:2006px}. Physically, the acceleration of black holes is caused by cosmic strings induced from conical singularities in the solution. We will take a typical solution to the vacuum Einstein equation in four dimensions into account, which describes a pair of rotating black holes constantly accelerating away from each other~\cite{Plebanski:1976gy, Hong:2004dm, Griffiths:2005se}, known as the accelerating Kerr black hole. Parallel to the stories in the Kerr$/$CFT, some aspects of the holographic properties for such solutions have been figured out. At extremality, the accelerating Kerr-Newman black hole acquires a warped and twisted product of AdS$_2\times$S$^2$ near horizon scaling region, and the asymptotic symmetry analysis can be applies to build a 2D CFT dual~\cite{Astorino:2016xiy}. For slowly accelerating Kerr black holes, the equation of motion for low frequency scalar perturbations in the near region still possesses hidden conformal symmetry~\cite{Siahaan:2018wvh}. The black hole entropy in the accelerating case still matches the result from the Cardy formula. See~\cite{Arenas-Henriquez:2023hur, Cisterna:2023qhh} for holographic properties of the accelerating black holes in lower dimensions. One next step is to ask whether the 2D CFT is unique for holographic descriptions of a generic accelerating Kerr black hole. Following the studying in the non-accelerating case~\cite{Aggarwal:2019iay}, we take a warped CFT as the possible dual field theory to the accelerating Kerr black hole.

The warped CFT is a two dimensional non-relativistic quantum field theory with warped conformal symmetries featured by one Virasoro algebra plus one $U(1)$ Kac-Moody algebra~\cite{Hofman:2011zj, Detournay:2012pc}. It was initiated from the investigation of the holography of a large class of geometries with $SL(2, R)\times U(1)$ isometry. Many aspects of the properties of the warped CFT have been uncovered~\cite{Hofman:2014loa, Castro:2015uaa, Detournay:2015ysa, Castro:2015csg, Song:2016gtd, Song:2017czq, Jensen:2017tnb, Apolo:2018eky, Wen:2018mev, Apolo:2018oqv, Song:2019txa, Chen:2019xpb, Gao:2019vcc, Chen:2020juc, Apolo:2020bld, Apolo:2020qjm, Chen:2022fte}. Natural choices of the corresponding gravitational theories are the warped AdS$_3$ spacetimes, where the asymptotic symmetries can be represented by the Virasoro and Kac-Moody algebra~\cite{Anninos:2008fx, Compere:2008cv, Compere:2009zj, Blagojevic:2009ek, Anninos:2010pm, Anninos:2011vd, Henneaux:2011hv}. The near horizon geometry of an extreme Kerr black hole also endowed with $SL(2, R)\times U(1)$ isometry, which can be viewed as a quotient of warped AdS$_3$ with fixed polar angle~\cite{Guica:2008mu}. So the warped CFT is also relevant to the Kerr black hole. The Virasoro and Kac-Moody algebra with non-trivial central extensions have been built in the covariant phase space for a non-extreme Kerr black hole, and a Cardy like DHH entropy formula of the warped CFT~\cite{Detournay:2012pc} assists to recover the entropy of the Kerr black hole~\cite{Aggarwal:2019iay}. The scattering amplitudes of scalar and gravitational perturbations on a generic Kerr background also have dual warped CFT interpretations~\cite{Xu:2023jex}. In the present paper, we generalize the work in~\cite{Aggarwal:2019iay} to the accelerating Kerr black hole case. We will figure out a possible dual field theory as the warped CFT by explicitly carrying out the Virasoro and Kac-Moody algebra with central extensions, and show the matching between the entropy from DHH formula and the entropy of rotating black holes with acceleration.

This paper is organized as follows. Sec.\ref{sec2} reviews the accelerating Kerr black hole spacetime as well as the hidden conformal symmetry of the Klein-Gordon equation on it. The global part of the warped conformal symmetry is specified by imposing a constraint in the momentum space of the scalar perturbation. In Sec.\ref{sec3}, we consider the linearized covariant charges associate to a set of vector fields. These vector fields are chosen so that the induced infinitesimal coordinate transformations keep the scalar equation and the constraint invariant. And their Fourier zero modes coincide with the scaling generators of the conformal coordinates. The commutation relations of the vector fields form a Virasoro and Kac-Moody algebra without central extensions. In the covariant phase space, the Dirac brackets of the covariant charges induced from these vector fields form a warped conformal algebra with central extensions. Sec.\ref{sec4} is for the consistency check for the validity of the warped conformal symmetries by reproducing the entropy of the accelerating Kerr black hole from DHH formula. Sec.\ref{sec5} is for summary and discussion. In App.\ref{seca}, we construct another sector of the Virasoro algebra with central extension, which indicates the known possibility of a 2D CFT being relevant holographic dual to the accelerating Kerr black hole.

\section{Hidden Conformal Symmetry}\label{sec2}
The metric of the accelerating Kerr black hole in the Boyer-Lindquist coordinates takes the following form~\cite{Hong:2004dm, Griffiths:2005se}
\be\label{BLg}
ds^2=-\frac{Q}{\Omega^2\rho^2}\left(dt-a\sin^2\theta d\phi\right)^2+\frac{P\sin^2\theta}{\Omega^2\rho^2}\left((r^2+a^2)d\phi-adt\right)^2+\frac{\rho^2}{\Omega^2}\left(\frac{dr^2}{Q}+\frac{d\theta^2}{P}\right)\,,
\ee
where
\bea
\Omega&=&1-\alpha r\cos\theta\,,~~\rho^2=r^2+a^2\cos^2\theta\,,\nn\\
P&=&1-2\alpha M\cos\theta+\alpha^2a^2\cos^2\theta\,,\\
Q&=&(1-\alpha^2r^2)\Delta\,,~~\Delta=r^2-2Mr+a^2\,.\nn
\eea
The three parameters $M$, $a=J/M$, and $\alpha$ are the black hole mass, the angular momentum per unit mass and the acceleration of the black hole, respectively. This solution of the vacuum Einstein equation characterises a pair of rotating black holes uniformly accelerating away from each other under the action of string-like forces represented by conical singularities along the axis of symmetry. Near the north ($\theta=0$) and south ($\theta=\pi$) poles, there are deficit and excess polar angles which induce the conical singularities
\be
\lim_{\theta\to0}\frac{2\pi}{\sin\theta}\sqrt{\frac{g_{\phi\phi}}{g_{\theta\theta}}}=2\pi\Theta_+\,,~~~~\lim_{\theta\to\pi}\frac{2\pi}{\sin\theta}\sqrt{\frac{g_{\phi\phi}}{g_{\theta\theta}}}=2\pi\Theta_-\,,
\ee
where $\Theta_{\pm}=1\mp2\alpha M+\alpha^2a^2$. The black hole outer ($+$) and inner ($-$) horizons are located at
\be
r_{\pm}=M\pm\sqrt{M^2-a^2}\,.
\ee
At $r=1/\alpha$, there is an accelerating horizon, which lays far beyond the black hole horizons when the acceleration is small. The outer horizon area $A_{BH}$ and the surface gravity $\kappa_H$ determine the Bekenstein-Hawking entropy and the Hawking temperature of the black hole respectively in the following forms
\be\label{ST}
S_{BH}=\frac{A_{BH}}{4}=\frac{2\pi Mr_+}{1-\alpha^2r_+^2}\,,~~~~T_{H}=\frac{\kappa_H}{2\pi}=\frac{(r_+-r_-)(1-\alpha^2r_+^2)}{8\pi Mr_+}\,.
\ee

Now let us first review the massless scalar perturbation on the accelerating Kerr black hole background and the hidden conformal symmetry appearing in the corresponding solution space. Then we will specify the symmetry to a subalgebra which generates the global warped conformal symmetry by imposing a constraint in the momentum space of the scalar perturbation. Such constraint is also responsible for the holographic interpretations of the scalar scattering process in terms of a warped CFT. For a perturbative massless scalar field $\Phi$, its equation of motion is given by the Klein-Gordon equation
\be
\nabla^{\mu}\nabla_{\mu}\Phi=0\,.
\ee
On the accelerating Kerr black hole background, the scalar field equation allows separable solutions of the form~\cite{Bini:2008mzd}
\be
\Phi=(1-\alpha r\cos\theta)e^{-i\omega t+im\phi}R(r)S(\theta)\,,
\ee
where $\omega$ and $m$ are the frequency and angular momentum of the scalar field. With this ansatz, the Klein-Gordon equation can be separated into the radial and angular equations~\cite{Bini:2008mzd}
\be\label{KGr}
\begin{aligned}
&\frac{d}{dr}\left(Q\frac{dR(r)}{dr}\right)+\bigg[\frac{(2Mr_+\omega-am)^2}{(r-r_+)(r_+-r_-)(1-\alpha^2r^2)}-\frac{(2Mr_-\omega-am)^2}{(r-r_-)(r_+-r_-)(1-\alpha^2r^2)}\\
&+(r^2+2Mr+4M^2+a^2)\omega^2-2am\omega-2\alpha^2r(r-M)\bigg]R(r)=\mathcal{K}R(r)\,,
\end{aligned}
\ee
\be\label{KGt}
\begin{aligned}
&\frac{1}{\sin\theta}\frac{d}{d\theta}\left(\sin\theta\frac{dY(\theta)}{d\theta}\right)-\bigg[\frac{m^2}{\sin^2\theta}+a^2\omega^2\sin^2\theta-2am\omega-\alpha^2(M^2-a^2)\sin^2\theta\bigg]\frac{Y(\theta)}{P^2}\\
&=-\mathcal{K}\frac{Y(\theta)}{P}\,,
\end{aligned}
\ee
where $\mathcal{K}$ is the separation constant and $Y(\theta)=\sqrt{P}S(\theta)$. To show the hidden conformal symmetry, we assume the low frequency limit
\be
M\omega\ll1\,,
\ee
and consider the scalar field equations in the near region
\be
r\omega\ll1\,,
\ee
as usual on the Kerr background~\cite{Castro:2010fd}. In addition to the non-accelerating case, we also need to impose further restriction on the acceleration of the black hole as discussed in~\cite{Siahaan:2018wvh}. We consider the pair of black holes which are slowly accelerating away from each other such that the near region is far within the accelerating horizon
\be
\alpha^2r^2\ll1\,.
\ee
Given these assumptions, $\alpha^2r^2$ is well approximated by $\alpha^2r_+^2$ in the near region and low frequency limit. The radial equation \eqref{KGr} thus is simplified to have two regular singular points at horizons $r=r_{\pm}$, which takes the form
\be\label{KGrn}
\frac{d}{dr}\left(\Delta\frac{dR(r)}{dr}\right)+\left[\frac{(2Mr_+\omega-am)^2/k_+^2}{(r-r_+)(r_+-r_-)}-\frac{(2Mr_-\omega-am)^2/k_+^2}{(r-r_-)(r_+-r_-)}\right]R(r)=\mathcal{K'}R(r)\,,
\ee
where $k_+=1-\alpha^2r_+^2$ and $\mathcal{K}'=\mathcal{K}/k_+$. This is the hypergeometric equation with solutions transforming in the representation of $SL(2, R)$, which indicates the existence of hidden conformal symmetry in the solution space of the scalar field. To explicitly show the symmetry, it is convenient to invoke the conformal coordinates used in~\cite{Siahaan:2018wvh}\footnote{We do not remove the conical singularity at $\theta=\pi$, so there is no $\Theta_-$ factor present.}
\bea
\omega^+&=&\sqrt{\frac{r-r_+}{r-r_-}}e^{2\pi T_R\phi}\,,\nn\\
\omega^-&=&\sqrt{\frac{r-r_+}{r-r_-}}e^{2\pi T_L\phi-\frac{k_+}{2M}t}\,,\label{ccrd}\\
y&=&\sqrt{\frac{r_+-r_-}{r-r_-}}e^{\pi(T_L+T_R)\phi-\frac{k_+}{4M}t}\,,\nn
\eea
where
\be\label{trtl}
T_R=\frac{r_+-r_-}{4\pi a}k_+\,,~~~~T_L=\frac{r_++r_-}{4\pi a}k_+\,.
\ee
By using these conformal coordinates, one can define $SL(2, R)$ generators in the following way~\cite{Castro:2010fd}
\bea
H_+&=&i\frac{\p}{\p\omega^+}\,,\nn\\
H_0&=&i\left(\omega^+\frac{\p}{\p\omega^+}+\frac{y}{2}\frac{\p}{\p y}\right)\,,\label{cfgen}\\
H_-&=&i\left((\omega^+)^2\frac{\p}{\p\omega^+}+\omega^+y\frac{\p}{\p y}-y^2\frac{\p}{\p\omega^-}\right)\,.\nn
\eea
These vector fields satisfy the $SL(2, R)$ Lie bracket algebra
\be
[H_0, H_{\pm}]=\mp iH_{\pm}\,,~~~~[H_-, H_+]=-2iH_0\,,
\ee
With these vector fields and given the near region radial equation \eqref{KGrn}, one can show that the $\theta$ independent part of the scalar field $\Phi(t, r, \phi)=e^{-i\omega t+im\phi}R(r)$ satisfies the Casimir equation of the $SL(2, R)$
\be
\mathcal{H}^2\Phi(t, r, \phi)=\ell(\ell+1)\Phi(t, r, \phi)\,,
\ee
where
\bea
\mathcal{H}^2&=&-H_0^2+\frac{1}{2}(H_+H_-+H_-H_+)\nn\\
&=&\frac{1}{4}\left(y^2\frac{\p^2}{\p y^2}-y\frac{\p}{\p y}\right)+y^2\frac{\p^2}{\p\omega^+\p\omega^-}\,,\label{qCas}
\eea
is the $SL(2, R)$ quadratic Casimir. The eigenvalue $\mathcal{K}'$ of the Casimir operator is assigned to be $\ell(\ell+1)$. This intuitively indicates that the scalar field with low frequency in the near region of a slowly accelerating Kerr black hole forms the representation of the $SL(2, R)$ symmetry with chiral conformal weight $\ell$. There is another set of $SL(2, R)$ generators ($\bar{H}_{\pm}, \bar{H}_0$) which can be obtained by exchanging $\omega^+$ and $\omega^-$ in the expressions \eqref{cfgen}. Combined with the unbarred generators, these are known as the hidden conformal symmetries~\cite{Siahaan:2018wvh}.

However, in this paper, we are considering the warped conformal symmetries appearing in the near region. Thus not all the barred generators are legal under this consideration. In terms of the Boyer-Lindquist coordinates, the zero modes of the hidden conformal generators can be written as
\bea
H_0&=&\frac{i}{2\pi T_R}\frac{\p}{\p\phi}+2i\frac{M}{k_+}\frac{T_L}{T_R}\frac{\p}{\p t}\,,\\
\bar{H}_0&=&-2i\frac{M}{k_+}\frac{\p}{\p t}\,.
\eea
As the generator $\bar{H}_0$ only has time derivative operation, the eigenvalue of $\bar{H}_0$ hence is proportional to the frequency $\omega$ of the scalar field. In studying the bulk scattering processes holographically with a warped CFT, we should view the frequency $\omega$ of the scalar perturbation as a fixed constant instead of a variable conjugate to the time $t$. The evidence can be found in the context of WAdS$/$WCFT~\cite{Song:2017czq} and Kerr$/$WCFT~\cite{Xu:2023jex}. This condition in the momentum space of the scalar field set the eigenvalue of $\bar{H}_0$ to be a fixed constant. However, the generators $\bar{H}_{\pm}$ have non-trivial commutation relations to $\bar{H}_0$, i.e., $[\bar{H}_0, \bar{H}_{\pm}]=\mp i\bar{H}_{\pm}$. So the $\bar{H}_{\pm}$ operating on the scalar field will change the eigenvalue of $\bar{H}_0$ which is not expected from a warped CFT interpretation. In other words, taking the warped conformal symmetries in the near region into account, the allowed global symmetry generators are $H_{0, \pm}$ and $\bar{H}_0$. $H_{0, \pm}$ are generating the $SL(2, R)$ symmetry and $\bar{H}_0$ is responsible for the $U(1)$ symmetry of the warped CFT.

The global warped conformal symmetry $SL(2, R)\times U(1)$ is spontaneously broken to $U(1)\times U(1)$ by the periodic identification of the angular coordinate $\phi$, i.e., $\phi\sim\phi+2\pi$. Under this periodic identification, the conformal coordinates transform as
\be\label{wid}
\omega^+\sim e^{4\pi^2T_R}\omega^+\,,~~~~\omega^-\sim e^{4\pi^2T_L}\omega^-\,,~~~~y\sim e^{2\pi^2(T_L+T_R)}y\,.
\ee
Only $H_0$ and $\bar{H}_0$ are invariant under these transformation generated by the $U(1)\times U(1)$ subgroup element
\be
e^{-i4\pi^2T_RH_0-i4\pi^2T_L\bar{H}_0}\,.
\ee
In the next section, we will see that an explicit Virasoro algebra plus an $U(1)$ Kac-Moody algebra with central extensions can be formulated in the covariant phase space of a set of specific choice of vector fields. The zero modes of these vector fields are proportional to the unbroken $U(1)\times U(1)$ generators mentioned in this section. The Virasoro companioned $U(1)$ Kac-Moody algebra characterize the local symmetries of a warped CFT. So the results in the present paper will support the warped CFT possibly being the holographic dual quantum field description of the accelerating Kerr black hole.

\section{Warped Symmetries in the Accelerating Kerr Geometry}\label{sec3}
The near region $SL(2, R)\times U(1)$ warped conformal symmetry of the scalar field on the accelerating Kerr black hole background is spontaneously broken to $U(1)\times U(1)$ symmetry, which can be viewed as time translational symmetries along
\be\label{wcrd}
t^+=2\pi T_R\phi\,,~~~~t^-=\frac{k_+}{2M}t-2\pi T_L\phi\,,
\ee
with energy eigenfunction and eigenvalues taking the forms
\be
e^{-i\omega t+im\phi}=e^{-i\omega_Rt^+-i\omega_Lt^-}\,,~~~~\omega_R=\frac{2M^2\omega-am}{k_+\sqrt{M^2-a^2}}\,,~~~~\omega_L=\frac{2M\omega}{k_+}\,.
\ee
The corresponding Hamiltonians are
\be
H_R=i\frac{\p}{\p t^+}\,,~~~~H_L=i\frac{\p}{\p t^-}\,.
\ee
In terms of the conformal coordinates \eqref{ccrd}, the Hamiltonians coincide the scaling operators
\bea
H_R&=&i\left(\omega^+\frac{\p}{\p\omega^+}+\frac{y}{2}\frac{\p}{\p y}\right)=H_0\,,\\
H_L&=&-i\left(\omega^-\frac{\p}{\p\omega^-}+\frac{y}{2}\frac{\p}{\p y}\right)=-\bar{H}_0\,,
\eea
which are also related to the zero modes of the warped conformal generators.

Following the discussion in~\cite{Aggarwal:2019iay}, where the warped CFT local symmetries were presented in the Kerr spacetime, we consider a set of diffeomorphisms induced by the following vector fields
\bea
\zeta(\epsilon)&=&\epsilon(\omega^+)\frac{\p}{\p\omega^+}+\frac{\p\epsilon(\omega^+)}{\p\omega^+}\frac{y}{2}\frac{\p}{\p y}\,,\label{ln}\\
p(\hat{\epsilon})&=&\hat{\epsilon}(\omega^+)\left(\omega^-\frac{\p}{\p\omega^-}+\frac{y}{2}\frac{\p}{\p y}\right)\,,\label{pn}
\eea
with functions $\epsilon(\omega^+)$ and $\hat{\epsilon}(\omega^+)$. The form of these vector fields are chosen so that the induced infinitesimal coordinate transformations will keep the $SL(2, R)$ quadratic Casimir $\mathcal{H}^2$ and the generator $\bar{H}_0$ invariant. Infinitesimally, the vector fields $\zeta(\epsilon)$ and $p(\hat{\epsilon})$ lead to the following conformal coordinate transformations
\be
\omega^+\to\omega^++\epsilon(\omega^+),~~~~\omega^-\to\omega^-,~~~~y\to y+\frac{\epsilon'(\omega^+)}{2}y\,,
\ee
and
\be
\omega^+\to\omega^+,~~~~\omega^-\to\omega^-+\hat{\epsilon}(\omega^+)\omega^-,~~~~y\to y+\frac{\hat{\epsilon}(\omega^+)}{2}y\,.
\ee
The $SL(2, R)$ quadratic Casimir $\mathcal{H}^2$ \eqref{qCas} and $\bar{H}_0=i(\omega^-\p_{\omega^-}+(y/2)\p_y)$ are invariant up to the leading order under above transformations. This means that the vector fields keep the radial equation of the scalar field as well as the eigenvalue of the generator $\bar{H}_0$ invariant. The vector fields depend on functions of coordinates, which should be viewed as local symmetries represented from both the dynamics and kinematics of the scalar field. The corresponding conserved charges of these local symmetries and their commutation relations will be figured out in the remaining part of this paper. The form of the vector fields \eqref{ln} and \eqref{pn} are the same as in the Kerr case~\cite{Aggarwal:2019iay} due to the same expressions of $\mathcal{H}^2$ and $\bar{H}_0$. However, the physical charges will differ from the Kerr case because of the acceleration of the black hole.

The functions $\epsilon(\omega^+)$ and $\hat{\epsilon}(\omega^+)$ are chosen so that the vector fields are periodic under \eqref{wid}, which is equivalent to require
\be
\epsilon(e^{4\pi^2T_R}\omega^+)=e^{4\pi^2T_R}\epsilon(\omega^+)\,,~~~~\hat{\epsilon}(e^{4\pi^2T_R}\omega^+)=\hat{\epsilon}(\omega^+)\,.
\ee
To find out the algebras for the above vector fields, the typical choices of these functions are using their Fourier modes~\cite{Aggarwal:2019iay}
\be
\epsilon_m=2\pi T_R(\omega^+)^{1+\frac{im}{2\pi T_R}}\,,~~~~\hat{\epsilon}_n=(\omega^+)^{\frac{in}{2\pi T_R}}\,,
\ee
where $m$ and $n$ are integers. Define $\zeta_n\equiv\zeta(\epsilon_n)$ and $p_n\equiv p(\hat{\epsilon}_n)$, which are the warped conformal generators in the coordinate space, one can figure out the following corresponding commutation relations
\bea
i[\zeta_m, \zeta_n]&=&(m-n)\zeta_{m+n}\,,\nn\\
i[\zeta_m, p_n]&=&-np_{m+n}\,,\label{calg}\\
i[p_m, p_n]&=&0\,.\nn
\eea
This is the Virasoro Kac-Moody (VKM) algebra without central extensions. The zero modes of the VKM algebra are proportional to the zero modes of the warped conformal generators
\be
\zeta_0=-i2\pi T_RH_0\,,~~~~p_0=-i\bar{H}_0\,.
\ee

Next, we will calculate the linearized covariant charges associated to the diffeomorphisms induced by the vectors \eqref{ln} and \eqref{pn} acting on the event horizon. The covariant charges will implement symmetries associated with the diffeomorphisms on a phase space through the Dirac brackets. The corresponding Dirac brackets are the quantum version of the classical algebras determined by the vectors now with central extensions and the central charges reflect the underling quantum degree of freedoms. The variation of the covariant charge associated to a vector $\zeta$ in general relativity has two parts
\be
\delta\mathcal{Q}=\delta\mathcal{Q}_{IW}+\delta\mathcal{Q}_{WZ}\,.
\ee
The first part is the Iyer-Wald charge~\cite{Iyer:1994ys}
\be
\delta\mathcal{Q}_{IW}(\zeta, h; g)=\frac{1}{16\pi}\int_{\p\Sigma}*F_{IW}\,.
\ee
The co-dimension two surface of integration $\p\Sigma$ will be chosen as the horizon bifurcation surface $\Sigma_{\mathrm{bif}}$. The two-form field $F_{IW}$ has components given by
\be
(F_{IW})_{\mu\nu}=\frac{1}{2}\nabla_{\mu}\zeta_{\nu}h+\nabla_{\mu}h^{\sigma}_{~\nu}\zeta_{\sigma}+\nabla_{\sigma}\zeta_{\mu}h^{\sigma}_{~\nu}+\nabla_{\sigma}h^{\sigma}_{~\mu}\zeta_{\nu}-\nabla_{\mu}h\zeta_{\nu}-(\mu\leftrightarrow\nu)\,,
\ee
where $h^{\mu\nu}$ is the variation of the inverse metric $g^{\mu\nu}\to g^{\mu\nu}+h^{\mu\nu}$ and $h=g_{\mu\nu}h^{\mu\nu}$. The second part is the Wald-Zoupas charge~\cite{Wald:1999wa}
\be
\delta\mathcal{Q}_{WZ}(\zeta, h; g)=\frac{1}{16\pi}\int_{\p\Sigma}i_{\zeta}\cdot(*X)\,.
\ee
The one-form field $X$ is constructed from the background geometry and linear in $h^{\mu\nu}$ for the consistency conditions on the covariant charges. The determination of $X$ is a case by case formulation which depends on the background geometry under considering. For the black hole bifurcation surface case, the candidate choice of $X$ is given by~\cite{Haco:2018ske}
\be
X=2h^{\nu}_{~\mu}\Omega_{\nu}dx^{\mu}\,,
\ee
where $\Omega_{\mu}$ is the component of the H\'{a}\'{\j}i\v{c}ek one-form which measures the rotational velocity of the horizon
\be\label{Hof}
\Omega_{\mu}=q^{\nu}_{~\mu}n^{\sigma}\nabla_{\nu}l_{\sigma}\,.
\ee
Here $l^{\mu}\p_{\mu}$ and $n^{\mu}\p_{\mu}$ are the null normal vectors to the future and past horizons respectively, and satisfy $l^{\mu}n_{\mu}=-1$. The null parameter of the vectors are chosen so that they are invariant under the $2\pi$ identification of the coordinate $\phi$. $q_{\mu\nu}=g_{\mu\nu}+l_{\mu}n_{\nu}+n_{\mu}l_{\nu}$ is the induced metric on $\p\Sigma$. Given the integrability, the covariant charges will form an algebra with Dirac bracket
\be
\{\mathcal{Q}_{\zeta_n}, \mathcal{Q}_{\zeta_m}\}=\mathcal{Q}_{[\zeta_n, \zeta_m]}+K_{m, n}\,,
\ee
where the central extension term is given by~\cite{Compere:2018aar}
\be
K_{m, n}=\delta\mathcal{Q}(\zeta_n, \mathcal{L}_{\zeta_m}g; g)\,.
\ee

We will specify the charge algebras associate to the classical commutation relations \eqref{calg}. Parallel to the discussion in~\cite{Aggarwal:2019iay}, we define the charge variations associated to the vectors $\zeta_n$ and $p_n$ as
\be\label{LPc}
\delta L_n\equiv\delta\mathcal{Q}(\zeta_n, h; g)\,,~~~~\delta P_n\equiv\delta\mathcal{Q}(p_n, h; g)\,.
\ee
Assuming these charges are integrable and the Dirac brackets are well defined in the sense that the charge itself can be realized as an operator generating diffeomorphisms on a Hilbert space, the charge algebras can be divided into three sectors with possible central extensions. One is the Virasoro sector
\be
\{L_n, L_m\}=(m-n)L_{m+n}+K_{m, n}\,,~~~~K_{m, n}=\delta\mathcal{Q}(\zeta_n, \mathcal{L}_{\zeta_m}g; g)\,.
\ee
Next is the Kac-Moody sector
\be
\{P_n, P_m\}=k_{m, n}\,,~~~~k_{m, n}=\delta\mathcal{Q}(p_n, \mathcal{L}_{p_m}g; g)\,.
\ee
Finally there is possible central term in the mixed sector
\be
\{L_n, P_m\}=mP_{m+n}+\mathfrak{R}_{m, n}\,,~~~~\mathfrak{R}_{m, n}=\delta\mathcal{Q}(\zeta_n, \mathcal{L}_{p_m}g; g)\,.
\ee
To figure out the central terms, we will use the conformal coordinates \eqref{ccrd}. The surface of integration $\p\Sigma$ is the black hole horizon bifurcation surface $\Sigma_{\mathrm{bif}}$, which is the intersection of the future and past horizons. In the original Boyer-Lindquist coordinates, the future horizon is located at $r=r_+, t\in(0, \infty)$ and the past horizon is located at $r=r_+, t\in(-\infty, 0)$. These are mapped to $\omega^-=0$ and $\omega^+=0$, respectively. So the bifurcation surface is at $\omega^+=\omega^-=0$. Near this co-dimension two surface, in terms of the conformal coordinates, the black hole metric \eqref{BLg} can be expanded as
\bea\label{cg}
ds^2&=&\frac{F_{+-}}{y^2}d\omega^+d\omega^-+\frac{F_{yy}}{y^2}dy^2+F_{\theta\theta}d\theta^2+\omega^+\frac{F_{-y}}{y^3}d\omega^-dy+\omega^-\frac{F_{+y}}{y^3}d\omega^+dy\nn\\
&+&\mathcal{O}((\omega^+)^2, \omega^+\omega^-, (\omega^-)^2)\,,
\eea
where
\bea
F_{+-}&=&\frac{4(r_+^2+a^2\cos^2\theta)}{(1-\alpha r_+\cos\theta)^2k_+}\,,\nn\\
F_{yy}&=&\frac{16J^2P\sin^2\theta}{(r_+^2+a^2\cos^2\theta)(1-\alpha r_+\cos\theta)^2k_+^2}\,,\nn\\
F_{\theta\theta}&=&\frac{r_+^2+a^2\cos^2\theta}{(1-\alpha r_+\cos\theta)^2P}\,,\label{FF}\\
F_{-y}&=&\frac{-2(8\pi J)^2T_R(T_L+T_R)+32\pi Ja^2T_R(k_+-P)\sin^2\theta}{(r_+^2+a^2\cos^2\theta)(1-\alpha r_+\cos\theta)^2k_+^3}\,,\nn\\
F_{+y}&=&\frac{-8(4\pi J)^2T_L(T_L+T_R)+8\left[4J^2+4\pi Ja^2(T_L+T_R)+a^2(r_+^2+a^2\cos^2\theta)\right]\sin^2\theta}{(r_+^2+a^2\cos^2\theta)(1-\alpha r_+\cos\theta)^2k_+^3}\nn\\
&+&\frac{64\pi^2a^4\left[T_L^2(2k_+P+2k_+^2-k_+-3)+T_LT_R(k_+P+k_+^2-k_+-1)\right]\sin^2\theta}{(r_+^2+a^2\cos^2\theta)(1-\alpha r_+\cos\theta)^2k_+^5}\nn\\
&-&\frac{8a^4k_+^2(k_+^2-1)\sin^4\theta}{(r_+^2+a^2\cos^2\theta)(1-\alpha r_+\cos\theta)^2k_+^5}\,.\nn
\eea
The null normal co-vector and vector of the future and past horizons are chosen respectively as
\be
l_{\mu}dx^{\mu}=y^{\frac{-2T_L}{T_L+T_R}}d\omega^-\,,~~~~n^{\mu}\p_{\mu}=-y^{\frac{2T_L}{T_L+T_R}}\p_{\omega^-}\,.
\ee
Note that the $y$ dependent factors are set for the periodicity under identifications \eqref{wid}. An observation made in~\cite{Haco:2018ske} is that the integrations on the bifurcation surface with metric \eqref{cg} have non-vanishing contributions coming from the simple poles in $\omega^+$ and the relevant integral has a part of the contribution
\be
\int_{\omega_0^+}^{e^{4\pi^2T_R}\omega_0^+}(\omega^+)^{-1+\frac{i(m+n)}{2\pi T_R}}d\omega^+=4\pi^2T_R\delta_{m, -n}\,.
\ee
The integral is performed near the reference point $\omega_0^+$ when $\omega^+\to0$. Taken this into account and using the functions \eqref{FF}, the variation of the charge along $\zeta_m$ on the bifurcation surface associated to $\zeta_n$ receives the following two contributions
\bea
\delta\mathcal{Q}_{IW}(\zeta_n, \mathcal{L}_{\zeta_m}g; g)&=&i\left((4\pi^2T_R^2)m+m^3\right)\delta_{m, -n}\nn\\
&\times&\int_0^{\pi}d\theta\frac{\sqrt{F_{yy}F_{\theta\theta}}(2F_{+-}+F_{-y}-F_{+y})}{16F_{+-}}\nn\\
&=&2i\frac{J}{k_+^2}\frac{T_R}{T_L+T_R}\left((4\pi^2T_R^2)m+m^3\right)\delta_{m, -n}\,,
\eea
\bea
\delta\mathcal{Q}_{WZ}(\zeta_n, \mathcal{L}_{\zeta_m}g; g)&=&i\left((4\pi^2T_R^2)m+m^3\right)\delta_{m, -n}\nn\\
&\times&\int_0^{\pi}d\theta\frac{\sqrt{F_{yy}F_{\theta\theta}}\left[2(T_L-T_R)F_{+-}-(T_L+T_R)(F_{-y}-F_{+y})\right]}{32(T_L+T_R)F_{+-}}\nn\\
&=&i\frac{J}{k_+^2}\frac{T_L-T_R}{T_L+T_R}\left((4\pi^2T_R^2)m+m^3\right)\delta_{m, -n}\,.
\eea
The central extension term in the Virasoro sector is determined by adding up the above two parts. The linear term in $m$ can be absorbed into the zero mode of the generators and we are left with
\be
K_{m, n}=i\frac{J}{k_+^2}m^3\delta_{m -n}\,.
\ee
The variation of the charge along $p_m$ associated to $p_n$ contains the following two parts
\bea
\delta\mathcal{Q}_{IW}(p_n, \mathcal{L}_{p_m}g; g)&=&-im\delta_{m, -n}\nn\\
&\times&\int_0^{\pi}d\theta\frac{\sqrt{F_{yy}F_{\theta\theta}}(2F_{+-}-F_{-y}+F_{+y})}{16F_{+-}}\nn\\
&=&-2i\frac{J}{k_+^2}\frac{T_L}{T_L+T_R}m\delta_{m, -n}\,,
\eea
\bea
\delta\mathcal{Q}_{WZ}(p_n, \mathcal{L}_{p_m}g; g)&=&im\delta_{m, -n}\nn\\
&\times&\int_0^{\pi}d\theta\frac{\sqrt{F_{yy}F_{\theta\theta}}\left[2(T_L-T_R)F_{+-}-(T_L+T_R)(F_{-y}-F_{+y})\right]}{32(T_L+T_R)F_{+-}}\nn\\
&=&i\frac{J}{k_+^2}\frac{T_L-T_R}{T_L+T_R}m\delta_{m, -n}\,.
\eea
Adding up these two parts gives the central extension term in the Kac-Moody sector
\be
k_{m, n}=-i\frac{J}{k_+^2}m\delta_{m, -n}\,.
\ee
Finally, the two parts of the charge variation along $p_m$ associated to $\zeta_n$ take the forms
\bea
\delta\mathcal{Q}_{IW}(\zeta_n, \mathcal{L}_{p_m}g; g)&=&\left((2\pi iT_R)m+m^2\right)\delta_{m, -n}\nn\\
&\times&\int_0^{\pi}d\theta\frac{\sqrt{F_{yy}F_{\theta\theta}}(F_{-y}-F_{+y})}{16F_{+-}}\nn\\
&=&-\frac{J}{k_+^2}\frac{T_L-T_R}{T_L+T_R}\left((2\pi iT_R)m+m^2\right)\delta_{m, -n}\,,
\eea
\bea
\delta\mathcal{Q}_{WZ}(\zeta_n, \mathcal{L}_{p_m}g; g)&=&i\left((2\pi iT_R)m+m^2\right)\delta_{m, -n}\nn\\
&\times&\int_0^{\pi}d\theta\frac{\sqrt{F_{yy}F_{\theta\theta}}\left[2(T_L-T_R)F_{+-}-(T_L+T_R)(F_{-y}-F_{+y})\right]}{32(T_L+T_R)F_{+-}}\nn\\
&=&\frac{J}{k_+^2}\frac{T_L-T_R}{T_L+T_R}\left((2\pi iT_R)m+m^2\right)\delta_{m, -n}\,.
\eea
These two parts cancels with each other which gives vanishing central term in the mixed sector
\be
\mathfrak{R}_{m, n}=0\,.
\ee
In deriving the central terms, the Wald-Zoupas counter-term with the H\'{a}\'{\j}i\v{c}ek one-form \eqref{Hof} is essential. There are two main motivations for this choice of the counter-term. One is that the central terms derived solely from the Iyer-Wald charge depend on the temperature of the dual warped CFT. The Wald-Zoupas counter-term makes the central terms temperature independent. Second is that it can eliminate the mixed central term from $L_n$ and $P_m$ bracket which is also expected from the canonical form of the warped conformal algebra. Putting the three sectors together, the charge algebra of \eqref{LPc} induced from \eqref{calg} can be expressed as
\bea
\{L_n, L_m\}&=&(m-n)L_{m+n}+i\frac{J}{k_+^2}m^3\delta_{m, -n}\,,\nn\\
\{L_n, P_m\}&=&mP_{m+n}\,,\label{wcalg}\\
\{P_n, P_m\}&=&-i\frac{J}{k_+^2}m\delta_{m, -n}\,.\nn
\eea
This is known as the warped conformal algebra consists of a Virasoro algebra and a $U(1)$ Kac-Moody algebra with the central charge $c$ and the Kac-Moody level $k$ being
\be\label{ck}
c=12\frac{J}{k_+^2}\,,~~~~k=-2\frac{J}{k_+^2}\,.
\ee
The point to emphasize here is that although the conformal coordinates \eqref{ccrd} are introduced to make the global warped conformal symmetry manifest when the acceleration parameter $\alpha$ is small, the charge algebra presented in \eqref{wcalg} is exact for $\alpha$. The conformal coordinates are valid for finite $\alpha$ in the near horizon region where $\alpha^2r^2$ is well approximated by $\alpha^2r_+^2$ for any $\alpha$ in the radial equation of the scalar field. In calculating the variation of the charges in the near horizon region, we perform a coordinate transformation from the Boyer-Lindquist type to the conformal one and allow $\alpha$ to take any value smaller than $1/r_+$. The presence of the warped conformal algebra suggesting that the accelerating Kerr black hole could be described holographically in terms of a warped CFT.

\section{Black Hole Entropy from Warped Symmetries}\label{sec4}
In this section, we will use the entropy formula derived from the modular property of the warped CFT to recover the entropy of the accelerating Kerr black hole in \eqref{ST}. We choose the coordinates $t^+$ and $t^-$ defined in \eqref{wcrd} as the finite temperature warped CFT coordinates. These coordinates are defined on a torus due to the Boyer-Lindquist coordinates $t$ and $\phi$ having the following spatial and thermal identifications
\be
(t, \phi)\sim(t, \phi+2\pi)\sim(t+i\beta, \phi+i\beta\gamma_H)\,,
\ee
where $\beta=1/T_H$ and $\gamma_H=a/(2Mr_+)$ are the inverse Hawking temperature and angular velocity for the event horizon. So the warped CFT coordinates inherit the following spatial and thermal identifications
\be\label{stid}
(t^+, t^-)\sim(t^++4\pi^2T_R, t^--4\pi^2T_L)\sim(t^++2\pi i, t^-+2\pi i)\,.
\ee
The above identifications determines a generic torus where the warped CFT lives on. To invoke the entropy formula, one can use the modular property of the warped CFT which manifests as a symmetry in the partition function for swapping the canonical spatial and thermal circles. Such a torus with canonical spatial circle can be obtained by a following warped conformal transformation
\be\label{htpm}
\hat{t}^+=-\frac{t^+}{2\pi T_R}\,,~~~~\hat{t}^-=t^-+\frac{T_L}{T_R}t^+\,,
\ee
after which the torus \eqref{stid} becomes a canonical one
\be\label{cstid}
(\hat{t}^+, \hat{t}^-)\sim(\hat{t}^+-2\pi, \hat{t}^-)\sim(\hat{t}^+-2\pi\hat{\tau}, \hat{t}^-+2\pi\hat{\bar{\tau}})\,,
\ee
where
\be\label{httb}
\hat{\tau}=\frac{i}{2\pi T_R}\,,~~~~\hat{\bar{\tau}}=\frac{i(T_L+T_R)}{T_R}\,.
\ee
In terms of the canonical torus \eqref{cstid}, the thermal entropy of the warped CFT takes the form~\cite{Aggarwal:2019iay, Detournay:2012pc}
\be\label{tS}
S_{(0|1)}(\hat{\bar{\tau}}|\hat{\tau})=2\pi i\frac{\hat{\bar{\tau}}}{\hat{\tau}}\hat{P}_0^{vac}+4\pi i\frac{1}{\hat{\tau}}\hat{L}_0^{vac}\,,
\ee
where $\hat{L}_0^{vac}$ and $\hat{P}_0^{vac}$ are the vacuum expectation values of the Virasoro and Kac-Moody zero modes on the canonical torus, respectively. These vacuum values are not independent since the vacuum state is parameterized by a spectral flowing in the warped CFT, and they are related through~\cite{Detournay:2012pc}
\be\label{LPrela}
\hat{L}_0^{vac}=-\frac{c}{24}+\frac{(\hat{P}_0^{vac})^2}{k}\,,
\ee
where $c$ is the Virasoro central charge and $k$ is the Kac-Moody level. On the canonical torus, the zero mode $\hat{L}_0$ is the conserved charge associate to the vector $\p/\p\hat{t}^+$ which is the rotational Killing vector $-\p/\p\phi$ of the original metric \eqref{BLg} given \eqref{htpm} and \eqref{wcrd}. So this charge is proportional to the angular momentum of the rotating black hole and thus has vanishing vacuum value
\be\label{L=0}
\hat{L}_0^{vac}=0\,.
\ee
Substituting the result \eqref{L=0} and the central extension parameters \eqref{ck} into \eqref{LPrela}, one can get
\be
(\hat{P}_0^{vac})^2=-\frac{J^2}{k_+^4}\,.
\ee
Given the vacuum values of the zero modes, the entropy in \eqref{tS} with thermal parameters \eqref{httb} can be evaluated as
\be
S_{(0|1)}(\hat{\bar{\tau}}|\hat{\tau})=4\pi^2\frac{J}{k_+^2}(T_L+T_R)=\frac{2\pi Mr_+}{k_+}\,,
\ee
which is precisely the thermal entropy of the accelerating Kerr black hole expressed in \eqref{ST}.

\section{Summary and Discussion}\label{sec5}
In this paper, we investigate the possible holographic dual to the accelerating Kerr black hole as a warped CFT. The accelerating Kerr black holes are exact solutions to the vacuum Einstein equation in four dimensions. These solutions characterize a pair of rotating black holes uniformly accelerating away from each other caused by the cosmic strings from conical singularities. Similar to the Kerr black hole, the hidden conformal symmetry applied to low frequency scalar perturbations in the near region also exists when the pair of rotating black holes are slowly accelerating. By imposing the constant frequency condition to the scalar perturbation, the allowed symmetry generators are reduced to that of the global warped conformal symmetry. To specify the local warped conformal symmetry, we implement a set of vector fields whose commutation relations form the Virasoro and Kac-Moody algebra without central extensions. These vector fields are settled according to the forms used in the Kerr black hole case now with acceleration parameter. They are chosen not from the first principle but inspired by the invariance of the scalar radial equation as well as the eigenvalue of the $U(1)$ generator under the corresponding infinitesimal coordinate shifts. These vector fields should be viewed as local symmetries represented from both the dynamics and kinematics of the scalar perturbation. It is interesting to find out the consistent boundary conditions which support the diffeomorphisms induced from the chosen vector fields.

In the covariant phase space of the chosen vector fields, the linearized covariant charges form the warped conformal algebra consists of a Virasoro algebra and a $U(1)$ Kac-Moody algebra with central extensions under Dirac brackets. The variations of covariant charges are calculated by the Iyer-Wald charge mended by the Wald-Zoupas counter-term, which is introduced to make the cental terms temperature independent as well as vanishing mixed central term. Since the covariant charges are evaluated near the horizon, in spite of the global warped conformal symmetry applied to the scalar perturbation in the near region when the accelerating is slow, the warped conformal algebra with central terms obtained here is exact for the acceleration parameter of the black hole.

The construction of the warped conformal symmetries indicates that the warped CFT is possibly relevant for the holographic descriptions of the accelerating Kerr black hole. Invoking the modular properties of the warped CFT partition function, the thermal entropy of a finite temperature warped CFT can be determined by the modular parameters and the vacuum charges of the symmetry generators, which gives a Cardy like DHH entropy formula. This entropy formula can be used to recover the entropy of the accelerating Kerr black hole, given the central terms obtained from the covariant charges' algebra. This is the first consistency check for the possible warped CFT dual of a accelerating Kerr black hole.

From the warped CFT perspective, the acceleration of the bulk spectime must have some field theory impacts which distinguish an accelerating Kerr black hole form a single Kerr black hole. One of the clues can be found from the periods $\hat{\tau}$ and $\hat{\bar{\tau}}$ of the thermal circle of a warped CFT defined on a canonical tours \eqref{httb}. Given the temperatures $T_R$ and $T_L$ \eqref{trtl}, the canonical thermal periods satisfy the following relation
\be
\frac{\hat{\bar{\tau}}(\hat{\bar{\tau}}-2i)}{\hat{\tau}^2}=k_+^2\,.
\ee
The acceleration of the bulk spacetime makes the combination of the canonical thermal periods on the left hand side of the above equation less than 1. The holographic interpretation of the bulk acceleration depends on the fully construction of the dictionary between the accelerating Kerr black hole and the warped CFT. The cosmic strings which induce the conical defects around the north and south poles in the bulk geometry might introduce heavy states in the dual warped CFT. These are interesting future topics to explore. This paper presents a first step to the dictionary at a symmetry level.

Some interesting questions are waiting to be clarified. For example, are those warped symmetries responsible for the dynamics of the scalar or higher spin perturbations on the accelerating Kerr black hole background? For a slowly accelerating Kerr black hole, the radial equation \eqref{KGr} at large distance, which approaches to the accelerating horizon, can be approximately written as
\be
\frac{d}{dr}\left((r^2-r_{\alpha}^2)\frac{dR(r)}{dr}\right)+\left(\frac{\left((r_{\alpha}^2+a^2)\omega-am\right)^2}{r^2-r_{\alpha}^2}+\mathcal{K}+2\right)R(r)=0\,,
\ee
where $r_{\alpha}=1/\alpha$ is the accelerating horizon radius. This is a associated Legender equation which also allows hypergeometric solutions. Is this fact indicates the existing of conformal symmetries near the accelerating horizon? In App.\ref{seca}, we attach an alternative construction of another sector of the Virasoro algebra. So it seems that the underling symmetries of the rotating black holes depends highly on the choice of the diffeomorphisms taken into account. One need a first principle guidance from the consistent boundary conditions to pick out the allowed diffeomorphisms. We leave these topics for future research.

\appendix

\section{Another sector of the Virasoro algebra}\label{seca}
Consider the following vector field
\be
\bar{\zeta}_n=2\pi T_L(\omega^-)^{1+\frac{1n}{2\pi T_L}}\frac{\p}{\p\omega^-}+(2\pi T_L+in)(\omega^-)^{\frac{in}{2\pi T_L}}\frac{y}{2}\frac{\p}{\p y}\,,
\ee
where $n$ is an integer. The commutation relation for the vectors $\bar{\zeta}_n$ with different $n$ satisfies the Virasoro algebra
\be
i[\bar{\zeta}_m, \bar{\zeta}_n]=(m-n)\bar{\zeta}_{m+n}\,.
\ee
Define the charge variation associated to the vector $\bar{\zeta}_n$ as
\be\label{Lb}
\delta\bar{L}_n\equiv\delta\mathcal{Q}(\bar{\zeta}_n, h; g)\,.
\ee
Choosing the null normal co-vector and vector of the past and future horizons respectively as
\be
l_{\mu}dx^{\mu}=y^{\frac{-2T_R}{T_L+T_R}}d\omega^+\,,~~~~n^{\mu}\p_{\mu}=-y^{\frac{2T_R}{T_L+T_R}}\p_{\omega^+}\,,
\ee
one can find the charge algebra of \eqref{Lb} as the Virasoro algebra with central extension
\be
\{\bar{L}_n, \bar{L}_m\}=(m-n)\bar{L}_{m+n}+\bar{K}_{m, n}\,,
\ee
where the central term $\bar{K}_{m, n}$ is given by
\be
\bar{K}_{m, n}=\delta\mathcal{Q}(\bar{\zeta}_n, \mathcal{L}_{\bar{\zeta}_m}g; g)\,.
\ee
To evaluate the central term, one use the non-vanishing contributions from the simple poles in $\omega^-$ on the bifurcation surface. The relevant integral now is
\be
\int_{\omega_0^-}^{e^{4\pi^2T_L}\omega_0^-}(\omega^-)^{-1+\frac{i(m+n)}{2\pi T_L}}d\omega^-=4\pi^2T_L\delta_{m, -n}\,,
\ee
near the reference point $\omega_0^-$ when $\omega^-\to0$. The variation of the charge along $\bar{\zeta}_m$ on the bifurcation surface associated to $\bar{\zeta}_n$ has the following two parts
\bea
\delta\mathcal{Q}_{IW}(\bar{\zeta}_n, \mathcal{L}_{\bar{\zeta}_m}g; g)&=&i\left((4\pi^2T_L^2)m+m^3\right)\delta_{m, -n}\nn\\
&\times&\int_0^{\pi}d\theta\frac{\sqrt{F_{yy}F_{\theta\theta}}(2F_{+-}-F_{-y}+F_{+y})}{16F_{+-}}\nn\\
&=&2i\frac{J}{k_+^2}\frac{T_L}{T_L+T_R}\left((4\pi^2T_R^2)m+m^3\right)\delta_{m, -n}\,,
\eea
\bea
\delta\mathcal{Q}_{WZ}(\bar{\zeta}_n, \mathcal{L}_{\bar{\zeta}_m}g; g)&=&-i\left((4\pi^2T_L^2)m+m^3\right)\delta_{m, -n}\nn\\
&\times&\int_0^{\pi}d\theta\frac{\sqrt{F_{yy}F_{\theta\theta}}\left[2(T_L-T_R)F_{+-}-(T_L+T_R)(F_{-y}-F_{+y})\right]}{32(T_L+T_R)F_{+-}}\nn\\
&=&-i\frac{J}{k_+^2}\frac{T_L-T_R}{T_L+T_R}\left((4\pi^2T_R^2)m+m^3\right)\delta_{m, -n}\,.
\eea
Adding up the above two parts and absorbing the linear term in $m$ into zero mode of the generators, we get the central term
\be
\bar{K}_{m, n}=i\frac{J}{k_+^2}m^3\delta_{m, -n}\,.
\ee
This leads to another sector of the Virasoro algebra
\be
\{\bar{L}_n, \bar{L}_m\}=(m-n)\bar{L}_{m+n}+i\frac{J}{k_+^2}m^3\delta_{m -n}\,,
\ee
with the central charge $\bar{c}$ being
\be
\bar{c}=12\frac{J}{k_+^2}\,.
\ee

\section*{Acknowledgement}
This work is supported by the NSFC Grant No. 12105045.

\end{document}